\documentclass[aps,pra,reprint,tightenlines,superscriptaddress,floatfix]{revtex4-1}

\usepackage{amsmath,amsfonts,amssymb}
\usepackage{mathrsfs}
\usepackage{xparse}
\usepackage{physics}
\usepackage{booktabs}
\usepackage{graphicx}
\usepackage[utf8]{inputenc}
\usepackage[english]{babel}
\usepackage{hyperref}
\usepackage{romannum}
\AtBeginDocument{\pagenumbering{arabic}}
\newcommand{\Prtransition}{$^3$H$_4$ $\to$ $^1$D$_2$ }
\newcommand{\PR}{Pr$^{3+}$ }
\renewcommand{\mathbf}{\boldsymbol}
\DeclareDocumentCommand\term{mmg}{%
  {$^{#1}$#2%
  \IfNoValueF {#3} {$_{#3}$}%
  }%
}
\newcommand{\degree}{$^{\circ}$}
\makeatletter
\newcommand\etal{\emph{et al}\@ifnextchar.{}{.\ }}
\newcommand\eryso{Er$^{3+}$:Y$_2$SiO$_5$\@ifnextchar.{}{\@ifnextchar,{}{ }}}
\newcommand\pryso{Pr$^{3+}$:Y$_2$SiO$_5$\@ifnextchar.{}{\@ifnextchar,{}{ }}}
\newcommand\euyso{Eu$^{3+}$:Y$_2$SiO$_5$\@ifnextchar.{}{\@ifnextchar,{}{ }}}
\newcommand\yso{Y$_2$SiO$_5$\@ifnextchar.{}{\@ifnextchar,{}{ }}}

\makeatother

\begin{document}
\title{Stark control of solid-state quantum memory with spin-wave storage}

\author{Mohammed K.\ Alqedra}
\author{Sebastian P.\ Horvath}
\author{Adam Kinos}
\author{Andreas Walther}
\author{Stefan Kr\"oll}
\author{Lars Rippe}
\email{physics@rippe.se}
\affiliation{Department of Physics, Lund University, P.O. Box 118, SE-22100 Lund, Sweden.}

\date{\today}

\begin{abstract}
Quantum memories for quantum communication need to be able to store photons for an extended time and then to release them on demand. This can be achieved in atomic frequency comb ensemble-based quantum memories by control pulses that transfer the excitation to and from long-lived spin states. However, such pulses can give rise to coherent and incoherent noise due to their interaction with the memory ensemble. In this article, we experimentally demonstrate the ability to switch off the coherent noise from such control pulses during the echo emission in a spin-wave quantum memory, using the linear Stark effect in rare-earth-ion doped crystals. By applying an electric field pulse, the echo emission was coherently switched off prior to the first spin transfer pulse, and the stored data pulse was restored only when both an optical recall pulse and a re-phasing electrical pulse were applied, giving a high degree of control of both desired and undesired emissions. We estimate the effectiveness of this technique by turning off the free induction decay of a narrow ensemble of ions. This technique can thus improve the noise performance of spin-wave storage at the single photon level by quenching coherent optical radiation created by the strong control pulses. The method demonstrated here represents a proof-of-principle that the spin-wave storage scheme can be combined with Stark control. The combined scheme serves as an addition to the toolbox of techniques that can be used to realize a full version of a quantum repeater.

\end{abstract}

\maketitle

\section{Introduction}
Optical quantum memories are a key component for several quantum information applications. For example, they are used for synchronization of entanglement swapping in quantum repeaters, which enables long distance quantum communication \cite{Briegel1998,Kimble2008Jun,Wehner2018Oct}. They are also essential for signal synchronization in linear-optics-based quantum computing schemes \cite{Knill2001Jan}. The ability to store single-photons for long times and retrieve them on-demand are some of the key requirements for a practical quantum memories \cite{Lvovsky2009Dec,Tittel2010Feb}. 

Rare-earth ions are considered to be an attractive platform to realize quantum memories. This is primarily due to their excellent optical and spin coherence properties at cryogenic temperatures \cite{Zhong2015Jan}. Furthermore, the inhomogeneously broadened optical transition provides another resource that can be spectrally tailored and used to realize strong light-matter coupling \cite{,Nilsson2004,Sinclair2014Jul}. The atomic frequency comb (AFC) is one of the actively investigated quantum memory schemes used in rare-earth ions systems \cite{Afzelius2009May,deRiedmatten2008Dec,Afzelius2010Jan,Afzelius2010Aug,Sabooni2013Mar}. In the standard AFC scheme, light is stored as a collective optical excitation in an inhomogeneously broadened ensemble of ions. The ions are spectrally shaped into a series of narrow, highly absorbing peaks with a predefined frequency separation. The retrieval time of the stored excitation is predetermined by the frequency separation between the peaks. In order to enable on-demand retrieval, the scheme is combined with two bright control pulses to transfer the optical excitation to a spin level, and to recall it back to the optical level, on-demand, where it continues to rephase \cite{Afzelius2009May,Afzelius2010Jan,Jobez2014Aug}.
It is, however, challenging to realize spin-wave storage in the single photon regime due to the excessive optical noise created by emission from ions excited by the strong spin control pulses \cite{Gundogan2015Jun,Timoney2013Aug,Jobez2015Jun,Bonarota2014Aug}. This emission could be incoherent fluorescence from ions off-resonantly excited by the control pulses. It could also be coherent emission which can take the following forms: (\romannum{1}) free induction decay (FID) emitted due to resonant excitation of background ions on the spin control pulse transition, (\romannum{2}) undesired echo emission due to off-resonant excitation of the AFC ensemble by the control field \cite{Timoney2013Aug}. 

In this paper, we demonstrate how all coherent noise sources can be strongly suppressed, by combining electric field effects at the nano scale with spin-wave storage, which has a potential to improve the single photon storage performance. 
In a previous work, Stark control was combined with the standard AFC scheme to realize a noise-free and on-demand control without the need for spin transfer pulses \cite{Horvath2021May}. Furthermore, the Stark effect has been previously combined with photon echoes \cite{Meixner1992Sep,Wang1992May,Graf1997,Chaneliere2008Jun}, and with spin echoes \cite{Arcangeli2016Jun}. The Stark effect is also used in the CRIB memory scheme, where gradient electric fields are applied macroscopically along the light propagation axes to coherently control the collective emission from a narrow ensemble of ions \cite{Nilsson2005Mar,Alexander2006Feb,Kraus2006Feb,Lauritzen2011Jan}. The scheme presented here is based on using the linear Stark effect to split the ion ensemble within the nano scale into two electrically distinct ions classes that can be coherently controlled using electric field pulses. By applying an appropriate electric field pulse, coherent oscillations of the two ion classes are put 180$^{\circ}$ out of phase before applying the first spin control pulse. This will consequently suppress any coherent emission, including the photon echo. The echo emission will stay quenched also after applying the spin control pulses until a second electric field pulse is applied. This second electric field pulse puts the stored collective excitation back in phase, and simultaneously switches off any coherent processes initiated in the time between the two electric field pulses. In particular, coherent emission from the spin transfer pulses which otherwise might interfere with the signal recalled from the memory. The region between the two electric field pulses, where the coherent optical noise is suppressed, will therefore be referred to as the Vegas region \footnote{What happens in Vegas, stays in Vegas!}. To achieve spin-wave storage in the single photon regime with high signal-to-noise ratio, it is important to prevent everything created during this Vegas region from extending beyond it.

In addition, the electric field control provides more timing flexibility for applying the first spin control pulse without risking an echo re-emission during the spin transfer. The second electric pulse can also be tuned independently to delay the echo emission after the second spin control pulse, introducing one more degree of control to the spin-wave quantum memory scheme.

\section{Theory}
\label{sec:theory}
\begin{figure*}[tbh!]
    \centering
    \includegraphics[width=0.8\textwidth]{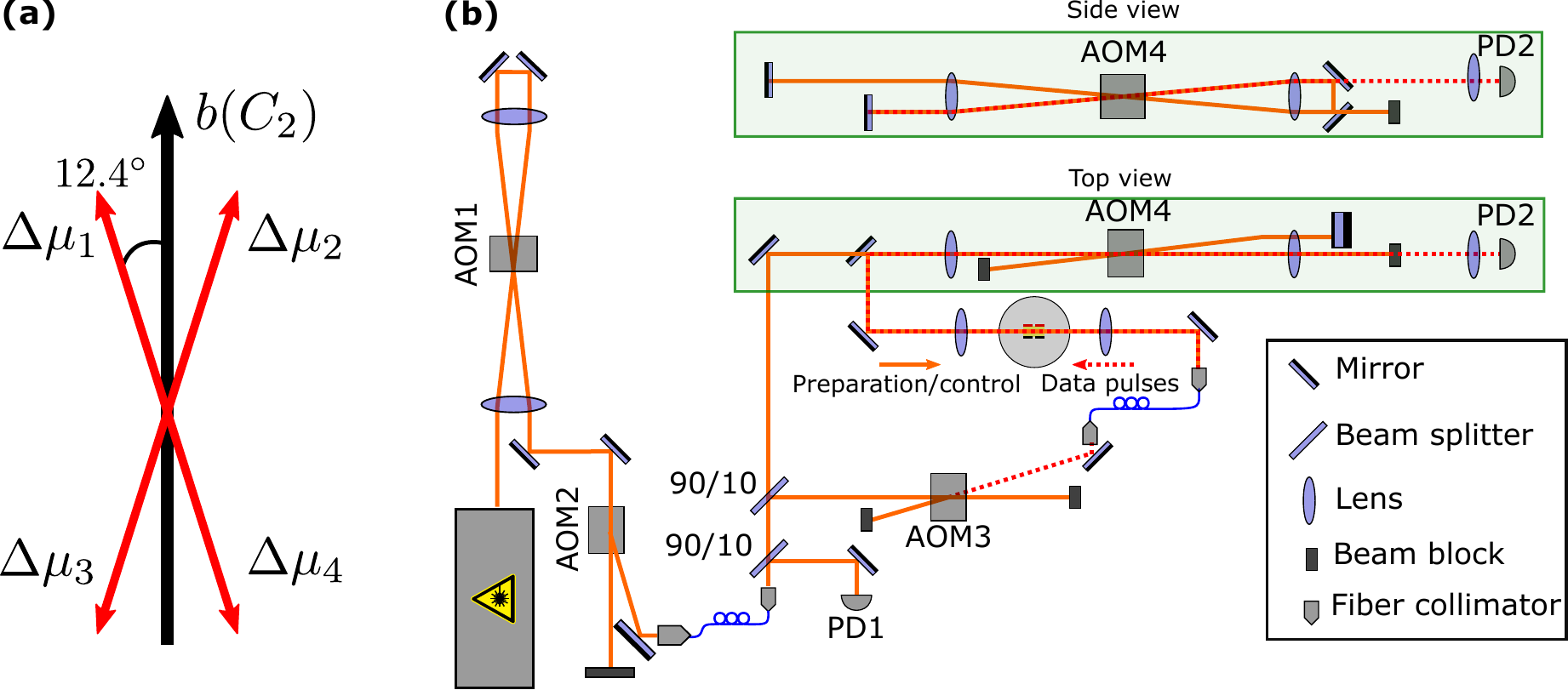}
    \caption{(a) The four possible static dipole moment orientations  for Pr$^{3+}$. (b) Experimental setup used for spin-wave storage. The orange solid line correspond to light used for AFC preparation and for spin control pulses, all propagating in the backward direction. The red dotted line represents the data pulses propagating in the forward direction. Both lights were overlapped within the crystal.}
    \label{fig:exp_setup}
\end{figure*}

Part of the theory discussed in this section was already presented in Ref. \cite{Horvath2021May}, and we repeat it here for convenience. It should be noted that although we only describe how the electric field can be used to control the phase evolution of the memory part here, the theory can be generalised to include all other parts that contribute to coherent emissions. The permanent electric dipole moment of the ground state differs from that of the excited state in Pr$^{3+}$. As a consequence, when an external electric field, \textbf{E}, is applied across the crystal, it will Stark shift the resonance frequency of the ions by a magnitude $\Omega$, which is given by:
\begin{equation}
    \Omega = \frac{\mathbf{\Delta\mu}\cdot\textrm{\textbf{E}}}{h},
\end{equation}

\noindent where $h$ is Planck's constant, and $\mathbf{\Delta\mu}$ is the difference in dipole moment between the ground state and the excited state. There are four possible orientations of $\mathbf{\Delta\mu}$ for \PR in \yso{}, all of them at an angle $\theta$ = 12.4\degree relative to the crystallographic $b$ axis as shown in Fig. \ref{fig:exp_setup} (a) \cite{Graf1997May}. For an electric field applied along the $b$ axis, the ions will split into two electrically distinct classes that experience the same magnitude of the Stark shift ($\Omega$), but with opposite signs. If the electric field is applied as a pulse with a finite duration, it will induce a phase shift of $+ \phi$ to one of the ion classes, and $- \phi$ to the other ion class, where $\phi$ is given by:
\begin{equation}
    \phi = 2\pi \int \Omega  dt.
\end{equation}

The spin-wave scheme is based on a three level configuration for storage. An incoming photon resonant with the optical transition $\ket{g}\to\ket{e}$ of the ions forming the AFC is stored as a collective optical excitation. In light of the distinction between the two electrically nonequivalent ions classes, the collective excitation can be described as \cite{Horvath2021May}:

\begin{equation}
	\ket{\psi(t)} = \frac{1}{\sqrt{2M}}\sum_{\ell = 0}^{M-1} e^{i \omega_{\ell} t} \left[e^{i \phi} \ket{\psi_{\ell}^+} + e^{-i \phi} \ket{\psi_{\ell}^-}\right].
	\label{eqn:col_ex}
\end{equation}
where $M$ the number of AFC peaks, $\omega_{\ell} = 2 \pi \Delta \ell$, with $\Delta$ being the spacing between the peaks. $\ket{\psi_{\ell}^\pm}$ are the wavefunctions of the positive and negative electrically inequivalent ion classes, which describe a delocalized optical excitation across the ions forming the peak, and is written as:

\begin{equation}
    \ket{\psi_{\ell}^\pm} = \frac{1}{\sqrt{N_{\ell}^\pm}}\sum_{j = 1}^{N_{\ell}^\pm} c_{\ell j}^\pm e^{2 \pi i \delta_{\ell j}^\pm t} e^{-i k z_{\ell j}^\pm}\ket{g_1 \ldots e_j \ldots g_{N_{\ell}^\pm}},	\label{eqn:st_wf}
\end{equation}

Here $N_{\ell}^\pm$ is the number of atoms in peak $\ell$ that experience a $\pm$ frequency shift due to \textbf{E}, $c_{\ell j}^\pm$ is the amplitude which depends on the spectral detuning $\delta_{\ell j}^\pm$ from the center of peak $\ell$, and on the position $z_{\ell j}^\pm$ of atom $j$ in AFC peak $\ell$, and $k$ is the photon wave vector.

The collective optical excitation described by Eq. \ref{eqn:col_ex} initially dephases due to the frequency separation of the AFC peaks, and rephases after times 1/$\Delta$ due to the periodicity of the AFC, leading to an echo emission. In the spin-wave scheme, a strong control pulse is applied before the echo emission to transfer the collective optical excitation to a spin level $\ket{s}$, converting it to a collective spin excitation where each term in the superposition state is written as $\ket{g_1 \ldots s_j \ldots g_{N_{\ell}^\pm}}$. This also freezes the dephasing due to $\omega_l$ and $\delta_{lj}$. A second strong control pulse is applied on-demand to reverse this process, after which the collective optical excitation continue to rephase and eventually emit an echo after a total storage time $T_s+1/\Delta$, with $T_s$ being the time spent in the spin state. 

By applying an electric field with $\int \Omega dt = 1/4$, before the first spin control pulse, the ions described by the two wavefunctions $\ket{\psi_{\ell}^\pm}$, given by Eq. \ref{eqn:st_wf}, will accumulate a $\pm\pi/2$ phase shift with respect to each other. As a result of this $\pi$ relative phase difference, the echo emission at 1/$\Delta$ will be turned off, giving more flexibility in the timing and the duration of the the first Spin control pulse without risking losing part of the echo due to rephasing during the spin transfer. After the second spin control pulse, the collective excitation will continue to evolve without echo emission until a second equivalent electric field pulse is applied. The second electric field pulse removes the $\pi$ relative phase difference between the two ion classes $\ket{\psi_{\ell}^\pm}$ in the AFC, and simultaneously adds a $\pi$ phase difference between the two classes of all other ions that were excited by the spin control pulses. This leads to an echo re-emission after $T_s+m/\Delta$, for m $\in \mathbb{N}$, and, at the same time, a suppression of all coherent background created within the Vegas region due to excitation by the spin control pulses.

\section{Experiment}
\label{sec:exp}
The experiment was performed on a 0.05\%-doped \pryso{}crystal cooled down to 2.1 K. The crystal had the dimensions of $6 \times 10 \times 10$ mm$^3$ for the $b \times D_1 \times D_2$ crystal axes, respectively. The top and bottom surfaces perpendicular to the $b$ axis were coated with gold electrodes, through which the electric field could be applied across the crystal. The AFC structure was prepared using the \Prtransition transition centered around 494.723 THz for \PR in site 1.

The optical setup used for the experiment is shown in Fig. \ref{fig:exp_setup}(b). The light source used was a frequency stabilised coherent 699-21 ring dye laser, tuned to the center of the inhomogeneous line of the \Prtransition in site 1, and polarized along the crystallographic $D_2$ axis. Light pulses were generated through a combination of the double-pass AOM1 and the single-pass AOM2 in series, through which the phase, frequency and amplitude of the pulses could be tailored. The light was the then split into two parts using 90:10 beam splitter, and the weaker beam was directly measured by the photodetector (PD1) as a reference to calibrate for laser intensity fluctuations. The rest of the light was split once more by another 90:10 beam splitter, and the weaker beam passed through the single pass AOM3, and propagated through the crystal in the forward direction (red dotted line). This light was used later to generate the data pulses to be stored. The stronger light that was transmitted through the beam splitter went through the double pass AOM4 setup, after which it propagated through the crystal in the backward direction (orange solid line). This light was used for the AFC preparation and for spin-transfer. The forward and backwards propagating beams were overlapped by maximizing the coupling of both beams through the two ends of the short fiber before the crystal.
\begin{figure}
    \centering
    \includegraphics[width=0.45\textwidth]{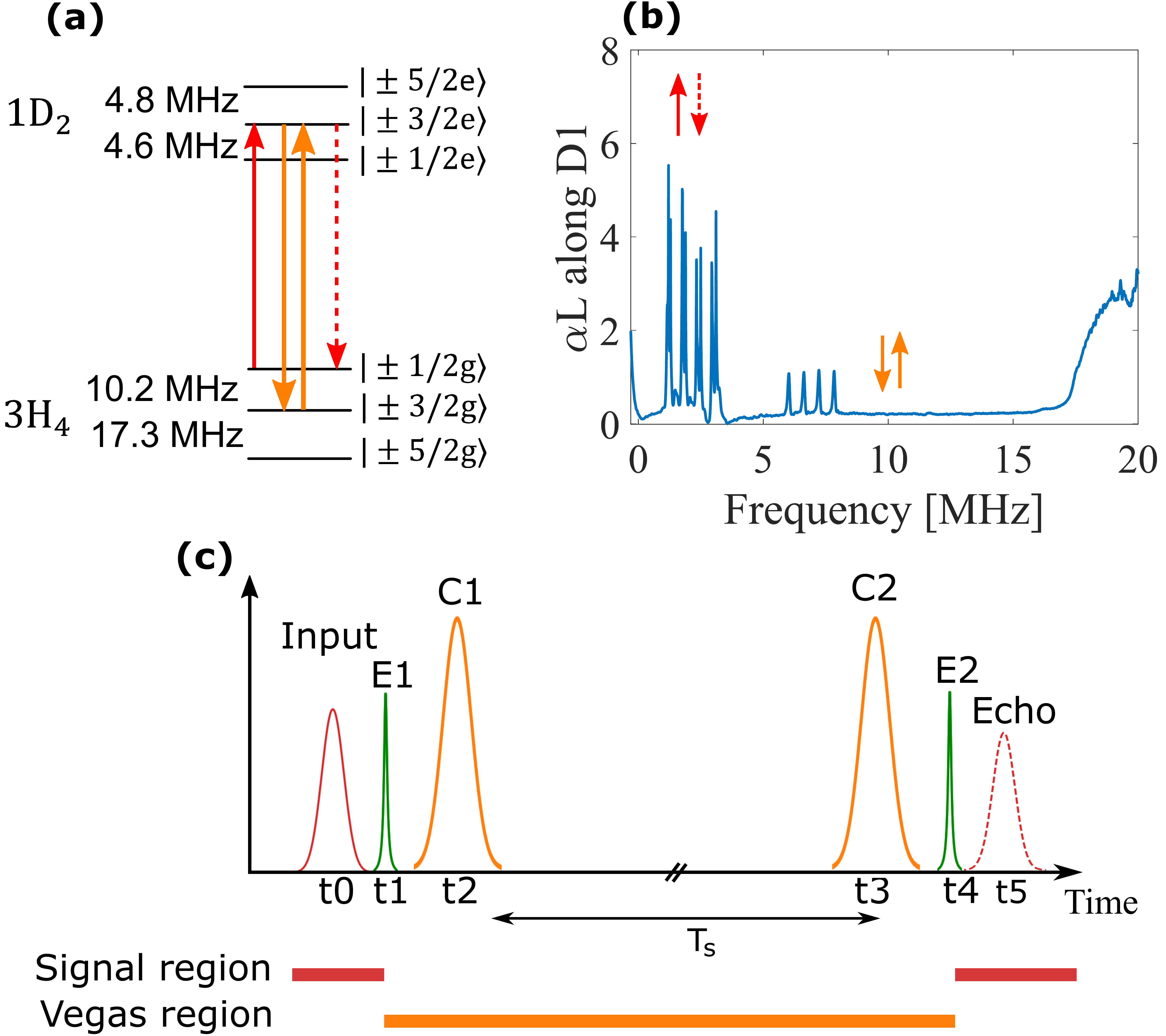}
    \caption{(a) level diagram of \PR showing the pulses used in the storage scheme. The solid red line is the input pulse, the two orange lines represent the 3 MHz FWHM control pulses used to transfer to and from the $\ket{3/2g}$ state, and the dashed red line is the echo. (b) A qualitative absorption measurement of the AFC structure structure, which also shows the spectral location of the pulses. The readout was distorted at the AFC peaks due to the high optical depth. The four peaks around 6.2 MHz correspond to the transition $\ket{1/2g}\to\ket{5/2e}$. (c) The pulse sequence in the time domain with signal and Vegas regions represented by the colored lines at the bottom. E1 and E2 are the first and second electric field pulses, C1 and C2 are the first and second spin control pulses. The lines in the bottom show which ions are affected by the electric field pulses at different times. E1 turns off coherent emission from ions excited at times $<$ t1, represented by the red line. E2 serves two purposes: it turns off all coherent background emission from ions excited between t1 and t4 by the two spin control pulses (orange line), and at the same time, E2 turns on emission that was turned off by E1 (red line), i.e., the stored photon echo. The gray line represents ions emitting incoherently, which is not affected by the electric field pulses.}
    \label{fig:pulse_seq}
\end{figure}
The memory output (propagating in the forward direction) passed through AOM4 in a single pass, as the AOM was turned off during the output. Furthermore, the output was spatially separated from the control beams and instead directed toward photodetector PD2. The crystal was mounted inside a bath cryostat cooling it down to 2.1 K. The light propagated through the crystal along the $D_1$ axis.
As mentioned earlier, the storage was performed on the \Prtransition transition in \PR. The level diagram of this transition is shown in Fig. \ref{fig:pulse_seq}(a). Both of the ground and the excited states have three hyperfine levels at zero magnetic field. In this experiment, the AFC is prepared in the $\ket{1/2g}$ level, the memory input is stored initially as an optical excitation in the $\ket{3/2e}$ level, and then transferred to the $\ket{3/2g}$ level as a spin-wave excitation.

Before preparing the AFC peaks, an 18 MHz wide transmission window was prepared in the center of the inhomogeneous line using the sequence described in Ref. \cite{Amari2010Sep}. The AFC was formed by coherently transferring back four narrow ensembles of ions to the $\ket{1/2g}$ level in the transmission window. This gave rise to four 140 kHz narrow absorption peaks separated by $\Delta =$ 600 kHz for their $\ket{1/2g}\to\ket{3/2e}$ transitions. As a result of those transfers, some unwanted ions were burned back to the $\ket{3/2g}$ level, and had to be cleaned away using frequency scan pulses with a scan range 8.5-14.5 MHz. The emptied $\ket{3/2g}$ level was used later for spin-wave storage. Due to the high absorption depth of the AFC peaks along the $D_2$ crystal axis, the weak frequency-scanned light used to probe the peaks was heavily distorted, which hindered a clean readout of the absorption structure. 
Therefore, the AFC preparation sequence was tested in a different \pryso{} crystal with a nominally equivalent praseodymium concentration, in which it was possible to have the light propagating along the $b$ crystal axis with a polarization along the less absorbing $D_1$ crystal axis. The crystal was 12 mm long along the $b$ axis. The absorption spectra measured along the $D_1$ crystal axis is shown in Fig. \ref{fig:pulse_seq}(b). Despite the lower absorption, the readout still has some distortions.

The pulse sequence used in the experiment is shown in Fig. \ref{fig:pulse_seq}(a)-(b) in frequency domain, and in Fig. \ref{fig:pulse_seq}(c) in the time domain with the signal and Vegas regions highlighted. A Gaussian pulse with 500 ns FWHM was used as a memory input. The pulse was resonant with the center of the AFC, which coincides with the $\ket{1/2g}\to\ket{3/2e}$ transition of the ensemble. At time t1, just after the input was absorbed (at t0), and well before the first control pulse (at t2), a Gaussian electric-field pulse, E1, with an amplitude of 54 V and FWHM = 23 ns was applied across the crystal through the gold-coated electrodes. This pulse introduced a relative phase shift of $\pm\pi/2$ for the two electrically inequivalent ion classes, which froze the echo re-emission. This allowed for considerable timing flexibility for the application of the first spin-transfer pulse without risking the echo being re-emitted during the transfer. The spin transfer was performed using 2 $\mu$s long complex hyperbolic secant (sechyp) pulses\cite{Rippe2005Jun,Roos2004Feb}. The first transfer pulse, C1, resonant with the $\ket{3/2e}\to\ket{3/2g}$, was applied at t2. It was used to transfer the collective excitation into the $\ket{3/2g}$ state, which froze the evolution of the atomic dipoles and convert the optical excitation to a spin-wave excitation. At t3, a second spin-transfer pulse, C2, was used to bring the spin state back to the excited state. The dipoles then continued to evolve as a collective excitation without emitting the echo, due to the $\pi$ phase difference introduced by the first electric field pulse. A second electric field pulse, E2, was then applied at t4 to remove the $\pi$ phase difference that was created by E1, and at the same time added a $\pi$ phase difference between the two classes of ions excited within the Vegas region by the spin transfer pulses, C1 and C2. This led to an echo emission at t5 and a suppression of coherent background created due to excitation by C1 and C2. The total storage time for this scheme is given by $\frac{m}{\Delta}+T_s$ for m $\in$ $\mathbb{N}$, with $T_s$ being the separation between C1 and C2. Here, in order to separate the echo emission from the second transfer pulse, the second electric field pulse was delayed such that the echo is emitted at the second re-phasing, i.e., using m = 2.

\section{Results and discussion}

\begin{figure*}[tbh!]
    \centering
    \includegraphics[width=0.73\textwidth]{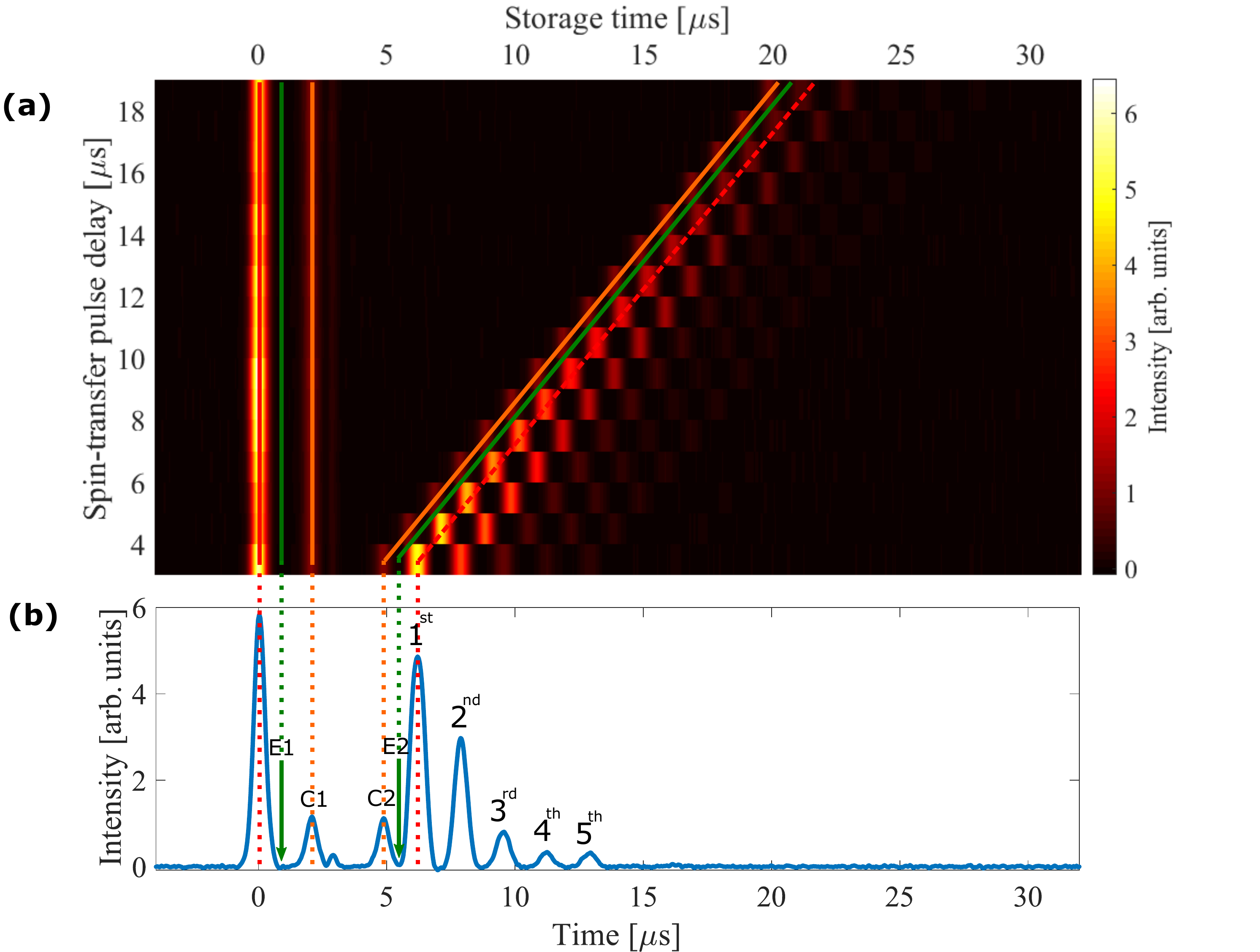}
    \caption{(a) Spin-wave storage with classical intensity input for varying storage times, and (b) for the shortest storage time. The first pulse marked by the solid red line is the part of the storage light transmitted through the AFC without being stored.  The first electric field pulse was applied directly after absorption at the time marked by the green line. The second faint pulse marked by the solid orange line is scattering of the first control pulse that leaked into the detection path. The second orange line (tilted) is scattering from the second control pulse applied at varying times. This is followed by another electric field pulse marked by the second green line (tilted). The dashed red lines is the restored echo after rephasing. The rest of the peaks peaks are higher order echoes emitted.}
    \label{fig:cmap}
\end{figure*}

The experiment was performed at varying storage times. The first electric and spin transfer pulses were fixed for all measurements. The second electric field pulse was delayed after the second control pulse such that the echo was emitted at the second re-phasing i.e. after $T_s + 2/\Delta$. This ensured that no part of the echo was emitted during the first or the second spin transfer pulses. By delaying the second electric field pulse, the recall of the echo can be further delayed after applying the second spin control pulse, which can be used as another degree of control. Here, both of the second spin control pulse and the second electric field pulse were delayed in steps of 1 $\mu$s to obtain different storage times. The result of this measurement is shown in Fig. \ref{fig:cmap}(a), with the storage sequence shown separately in Fig. \ref{fig:cmap}(b) for the shortest storage time. It is worth noting that the detection was performed in the forward direction, i.e. opposite to the control pulse propagation. Nevertheless, reflection of the control pulses, C1 and C2, from optical surfaces leaked into the detection path, and is marked by the two orange solid lines shown in Fig. \ref{fig:cmap}(a). There are several ways to reduce this reflection of the control pulses, for example using optical surfaces with anti-reflective coating, or by having a small angle between the control beam and the storage beam. The two green lines in the figure indicate the times when the two electric field pulses were applied. A recall of the second order echo as well as four other higher order echos can be seen in the figure. 

A challenge when implementing the spin-wave storage scheme at the single photon level is the optical noise created due to the control pulses, which can be either incoherent fluorescence or coherent collective emission such as FID and off-resonant echoes. When applying the second electric field to switch on the signal echo emission, it simultaneously shifts the phases of all ions contributing to the coherent optical noise, which consequently turns off the coherent noise contribution from these ions.
\begin{figure}[tbh!]
    \centering
    \includegraphics[width=0.45\textwidth]{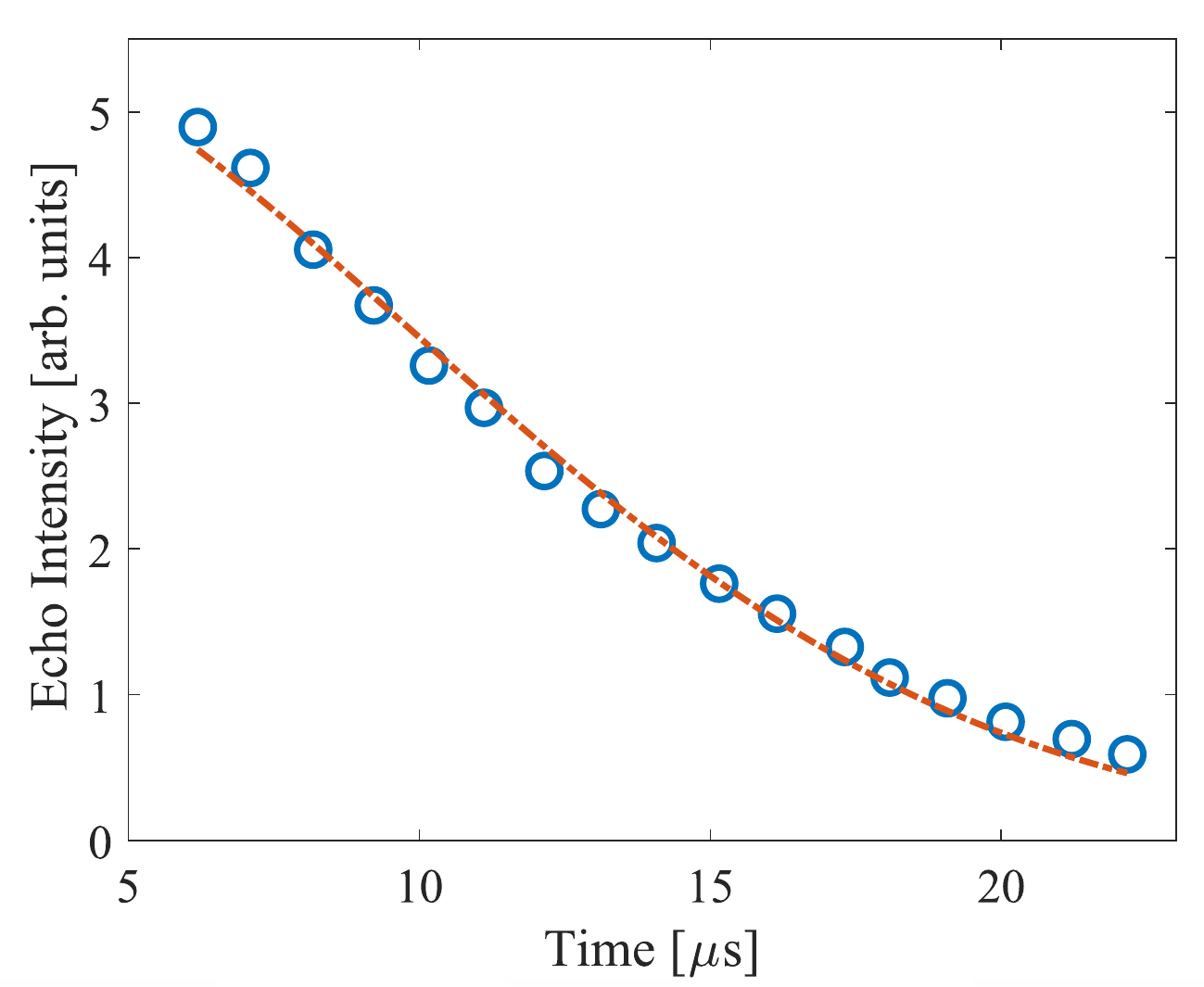}
    \caption{The circles are the maximum echo intensity at different spin storage time $T_s$, the dashed line is a Gaussian fit to Eq. \ref{eq:gauss}, giving an inhomogeneous spin linewidth of 26.8 $\pm$ 0.8 kHz}
    \label{fig:inh_decay}
\end{figure}

The exponential decay of the echo intensity, which can also be seen in Fig. \ref{fig:inh_decay}, is attributed to the inhomogeneous broadening of the spin transition. This was confirmed by fitting the echo intensity against the spin storage duration (T$_s$) to the following Gaussian \cite{Timoney2012Jun}:

\begin{equation}
    \textrm{I(T}_s\textrm{)} = \textrm{I}_0 \times \textrm{exp}\left[\frac{-(\gamma_{IS}T_s)^2}{2\textrm{ log}(2)/\pi^2} \right],
    \label{eq:gauss}
\end{equation}

where $I_0$ is a constant, and $\gamma_{IS}$ is the inhomogeneous spin linewidth. From the Gaussian fit, we obtain an inhomogeneous spin linewidth of 26.8 $\pm$ 0.8 kHz, in agreement with previous measurements in the same material\cite{Afzelius2010Jan,Gundogan2015Jun}. We see no contribution of any additional dephasing due to our phase switching technique using the electric field.
\begin{figure}[tbh!]
    \centering
    \includegraphics[width=0.45\textwidth]{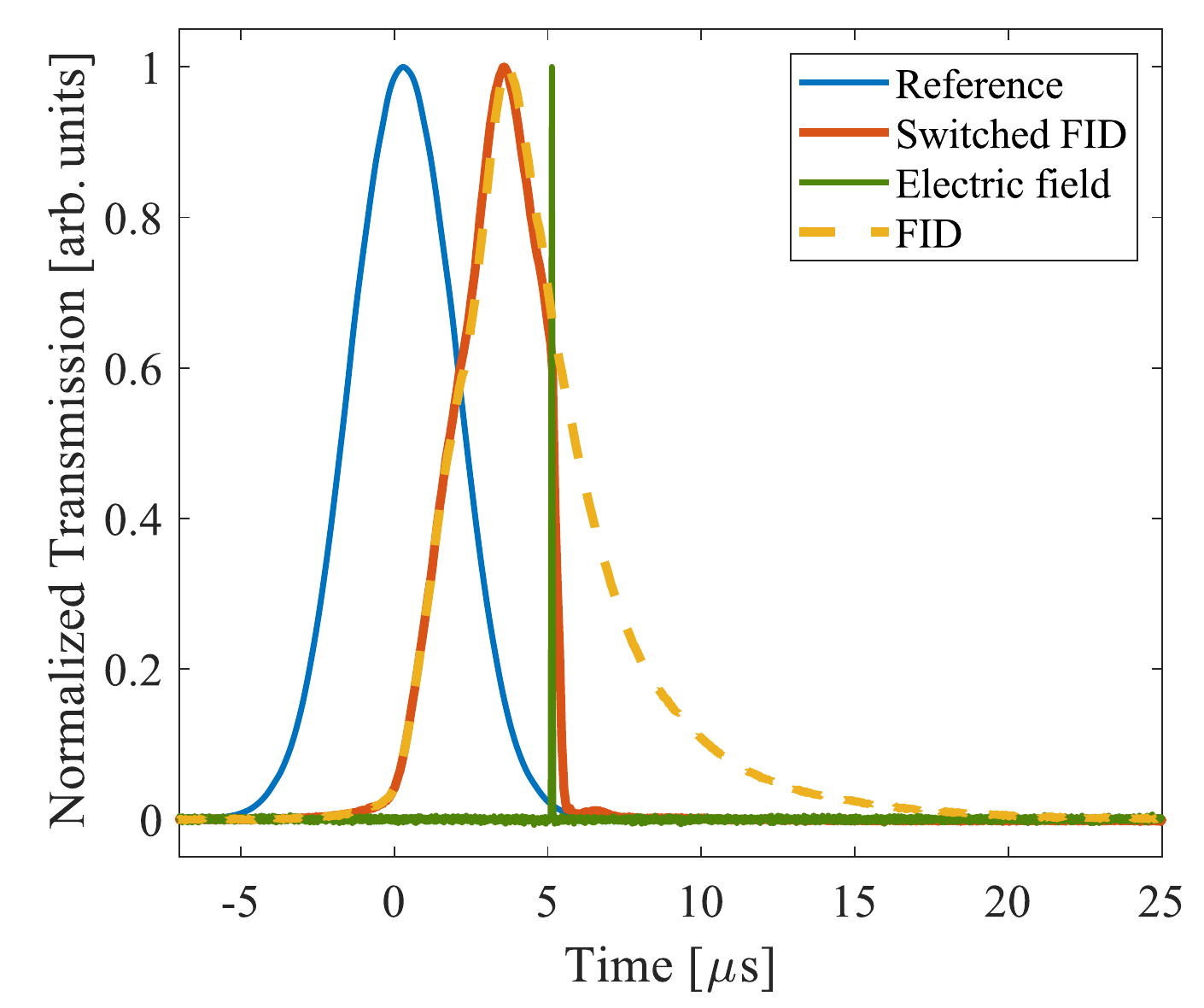}
    \caption{FID quenching using an electric field pulse. The blue line is the Gaussian pulse before propagating through the crystal. The yellow dashed line is the transmission after going through the narrow peak, with an extended FID emission. The delay on the transmission is caused by the slow light effect due to the strong dispersion across the transmission window. The green line shows the electric field pulse applied to switch off the FID emission. The red solid line shows the FID when it is turned off, also delayed due to dispersion.}
    \label{fig:switched_FID}
\end{figure}

\subsection{Suppression of free-induction decay}
In order to investigate the capacity of this technique to suppress coherent FID noise, a single 140 kHz peak was burned back in an empty 18 MHz wide transmission window. A Gaussian pulse with FWHM of 4 $\mu$s resonant with the peak was sent through the crystal. After absorbing the pulse, the ions in the narrow peak oscillate in phase and emit a coherent FID for a duration defined by the peak width. By applying an electric field pulse, the two electrically inequivalent ions classes were put out of phase, which consequently lead to a suppression of the FID emission. This was performed using classical light intensities to demonstrate the effect, and the result is shown in Fig. \ref{fig:switched_FID}. The integrated emission during the time interval $\left[6,25\right] \mu$s is reduced by a factor of $\sim$ 44 when the electric field pulse was applied compared to no pulse. In theory, the suppression of coherent emission is only limited by the electric field inhomogeneity across the light propagation path in the crystal. Although we only show the switching of the FID here, the same can be expected for all other kinds of coherent noise. This includes coherent off-resonantly excited echo emitted at the same frequency as the input, referred to as OREO in \cite{Timoney2013Aug}, and caused by off-resonant excitation of the comb structure by the spin control pulses. In addition, a two-pulse photon echo is another possible noise source  that can be switched off by this technique, which is created due to the two spin control pulses, and emitted at time $T_s$ after the second pulse. 

Since the presented technique only affects coherent optical noise, it would be most effective when used in materials where the majority of the optical noise is coherent, and thus can be quenched electrically without recourse to complicated spectral filtering.

It should be noted that the FID switching technique presented here is different from the one discussed in Ref. \cite{Minar2009Nov}, where an electric-field gradient dephases atoms on a macroscopic scale along the light propagation direction. In contrast, the control is achieved here by switching groups of ions out of phase on the submicron scale using discrete homogeneous electric field pulses. This is an essential difference, since microscopic cancellation may turn off the unwanted emissions in all directions.

\subsection{Fluorescence estimation}
Spin state storage at the single photon level is challenging and will be strongly affected by both coherent and incoherent fluorescence noise as mentioned earlier. Using the present technique, all sources of coherent noise can be reduced to such a high degree that fluorescence from off-resonant excitation will be the main limiting factor. The current experiments were performed with the \PR ion due to experimental availability, although it is not an ideal choice. Experiments at the single photon level demonstrated an optical noise level of $\sim$ 0.05 photon per shot within the echo emission duration. This is about an order of magnitude higher than the signal level expected from storage of a single photon. The optical noise was measured by applying the AFC sequence, described in Section \ref{sec:exp}, but without a storage pulse. The emission after the second electric pulse was collected with an optical collection efficiency of 40$\%$, and was detected using a Laser Components Count 50N avalanche photodiode with a quantum efficiency of 0.69 at 606 nm and a dark-count rate of 26 Hz. A Chroma band-pass filter (ET590/33m) was mounted before the detector to block light at wavelengths above 610 nm. 

To explore the potential of our technique, we here make an estimate of the fluorescence noise in other more suitable materials. In order to benchmark materials where the presented technique would be effective, we look into parameters that would lead to negligible incoherent fluorescence emission. 
The fluorescence is emitted due to off-resonant excitation of ions outside the spectral transmission window. To estimate the fluorescence noise, we look at the remaining absorption at the center of a spectral transmission window due to the Lorentzian tail of the ions outside. The total off-resonant absorption ($\alpha_c$) at the center of the transmission window can be written as \cite{Horvath2022Mar}:
\begin{equation}
    \alpha_c = \frac{2}{\pi}\frac{\Gamma_h}{\Delta}\alpha_0.
    \label{eq:alpha_c}
\end{equation}

Here $\Gamma_h$ is the homogeneous linewidth of the ions, $\Delta$ is the width of the transmission window, and $\alpha_0$ is the absorption outside the transmission window. For a crystal of length $L$, the power absorbed ($P_{abs}$) by the off resonant ions can be written as:

\begin{equation}
    P_{abs} = P_{in} [1 - e^{\alpha_c L}],
    \label{eq:P_abs}
\end{equation}

where $P_{in}$ is the input power. Equation \ref{eq:alpha_c} shows that materials with narrow homogeneous linewidth and in which wide spectral transmission windows can be created are favourable to reduce the off-resonant excitation. Here, we look into Eu$^{3+}$, which has an optical homogeneous linewidth of 122 Hz in site 1, an excited state lifetime of 1.9 ms \cite{Equall1994Apr}, and a branching ratio of $\sim$ 11\% to the $^7$F$_0$ zero phonon line, calculated from the lifetime of the excited stated and the dipole moment of the transition \cite{Hilborn1998Jul,Graf1998Sep}. For a 1\% doping concentration in \yso, the absorption depth of Eu$^{3+}$  along the $D_1$ crystal axis is 3.9 cm$^{-1}$ \cite{Konz2003Aug}. For a given input power, the part that will be off-resonantly absorbed by the ions outside a 40 MHz wide spectral transmission window in Eu$^{3+}$ is $\sim$ 20 ppm of the input. All off-resonantly absorbed photons are assumed to be re-emitted as fluorescence. Out of the total isotropic fluorescence, only the part that is overlapping with the spatial mode of the echo emission contributes to the optical noise. Here we assume that a diameter of 1 mm of the fluorescence light is collimated 20 cm after the crystal, which is only 1 ppm of the total isotropic emission. Furthermore, fluorescence photons are emitted at different times with some decay constant given by the excited state lifetime. Only photons emitted during the same time bin as the stored signal photon will contribute to the optical noise. For a 1 $\mu$s time bin at the beginning of the fluorescence decay, the probability of photon emission will be $\sim$ 0.1\%. Assuming a 1 $\mu$s long control pulse with 100 mW power, this will lead to an average of $\sim$ 10$^{-4}$ incoherent fluorescence photon emitted in a 1 $\mu$s time bin.
At such low level of incoherent noise, our presented technique can quench the other coherent noise emissions such as FID and off-resonant echoes, allowing for single photon storage without the need for additional spectral filtering. The lower optical depth in Eu$^{3+}$ can be compensated by a cavity to enhance the memory efficiency as has been demonstrated in Ref. \cite{Jobez2014Aug}. 
For comparison, doing the same calculation for the Pr$^{3+}$ ions used in the experiment, we estimate $\sim$ 0.03 incoherent fluorescence photon in a 1 $\mu$s time bin, taking into account experimental parameters, such as collection and detection efficiencies as well as control pulses with 10 mW power. This noise level is very close to the measured background mentioned earlier, and shows that our model gives reasonable predictions.\\ 

\section{Conclusion}
We used the linear Stark effect to coherently control the emission of the echo in the spin-wave storage scheme using electric field pulses. The first electric field pulse was used to switch off the echo emission after the absorption of the storage pulse, giving more time flexibility for applying the first spin control pulse. Then after the second spin control pulse, we used another electric field pulse to turn on the echo emission. We also showed that this technique can turn off the FID emission, and could therefore be used to quench the coherent optical emissions induced by the strong spin control pulses when performing the spin-wave storage at the single-photon level. If used in Eu$^{3+}$ :\yso, this technique has potential to enable spin-wave storage of single photons without the need for additional spectral filtering, which would substantially simplify noise-free quantum memory experiments.

\section{\label{sec:ack} ACKNOWLEDGMENTS}
This research was supported by the Swedish Research Council (no. 2016-05121, no. 2015-03989, no. 2016-04375, no. 2019-04949 and no. 2021-03755), the Knut and Alice Wallenberg Foundation (KAW 2016.0081), the Wallenberg Center for Quantum Technology (WACQT) funded by The Knut and Alice Wallenberg Foundation (KAW 2017.0449), and the European Union FETFLAG program, Grant No. 820391 (SQUARE) (2017.0449).

\bibliography{afcSW}

\begin{thebibliography}{41}%
\makeatletter
\providecommand \@ifxundefined [1]{%
 \@ifx{#1\undefined}
}%
\providecommand \@ifnum [1]{%
 \ifnum #1\expandafter \@firstoftwo
 \else \expandafter \@secondoftwo
 \fi
}%
\providecommand \@ifx [1]{%
 \ifx #1\expandafter \@firstoftwo
 \else \expandafter \@secondoftwo
 \fi
}%
\providecommand \natexlab [1]{#1}%
\providecommand \enquote  [1]{``#1''}%
\providecommand \bibnamefont  [1]{#1}%
\providecommand \bibfnamefont [1]{#1}%
\providecommand \citenamefont [1]{#1}%
\providecommand \href@noop [0]{\@secondoftwo}%
\providecommand \href [0]{\begingroup \@sanitize@url \@href}%
\providecommand \@href[1]{\@@startlink{#1}\@@href}%
\providecommand \@@href[1]{\endgroup#1\@@endlink}%
\providecommand \@sanitize@url [0]{\catcode `\\12\catcode `\$12\catcode
  `\&12\catcode `\#12\catcode `\^12\catcode `\_12\catcode `\%12\relax}%
\providecommand \@@startlink[1]{}%
\providecommand \@@endlink[0]{}%
\providecommand \url  [0]{\begingroup\@sanitize@url \@url }%
\providecommand \@url [1]{\endgroup\@href {#1}{\urlprefix }}%
\providecommand \urlprefix  [0]{URL }%
\providecommand \Eprint [0]{\href }%
\providecommand \doibase [0]{http://dx.doi.org/}%
\providecommand \selectlanguage [0]{\@gobble}%
\providecommand \bibinfo  [0]{\@secondoftwo}%
\providecommand \bibfield  [0]{\@secondoftwo}%
\providecommand \translation [1]{[#1]}%
\providecommand \BibitemOpen [0]{}%
\providecommand \bibitemStop [0]{}%
\providecommand \bibitemNoStop [0]{.\EOS\space}%
\providecommand \EOS [0]{\spacefactor3000\relax}%
\providecommand \BibitemShut  [1]{\csname bibitem#1\endcsname}%
\let\auto@bib@innerbib\@empty
%</preamble>
\bibitem [{\citenamefont {Briegel}\ \emph {et~al.}(1998)\citenamefont
  {Briegel}, \citenamefont {D\"ur}, \citenamefont {Cirac},\ and\ \citenamefont
  {Zoller}}]{Briegel1998}%
  \BibitemOpen
  \bibfield  {author} {\bibinfo {author} {\bibfnamefont {H.-J.}\ \bibnamefont
  {Briegel}}, \bibinfo {author} {\bibfnamefont {W.}~\bibnamefont {D\"ur}},
  \bibinfo {author} {\bibfnamefont {J.~I.}\ \bibnamefont {Cirac}}, \ and\
  \bibinfo {author} {\bibfnamefont {P.}~\bibnamefont {Zoller}},\ }\href
  {\doibase 10.1103/PhysRevLett.81.5932} {\bibfield  {journal} {\bibinfo
  {journal} {Phys. Rev. Lett.}\ }\textbf {\bibinfo {volume} {81}},\ \bibinfo
  {pages} {5932} (\bibinfo {year} {1998})}\BibitemShut {NoStop}%
\bibitem [{\citenamefont {Kimble}(2008)}]{Kimble2008Jun}%
  \BibitemOpen
  \bibfield  {author} {\bibinfo {author} {\bibfnamefont {H.~J.}\ \bibnamefont
  {Kimble}},\ }\href {\doibase 10.1038/nature07127} {\bibfield  {journal}
  {\bibinfo  {journal} {Nature}\ }\textbf {\bibinfo {volume} {453}},\ \bibinfo
  {pages} {1023} (\bibinfo {year} {2008})}\BibitemShut {NoStop}%
\bibitem [{\citenamefont {Wehner}\ \emph {et~al.}(2018)\citenamefont {Wehner},
  \citenamefont {Elkouss},\ and\ \citenamefont {Hanson}}]{Wehner2018Oct}%
  \BibitemOpen
  \bibfield  {author} {\bibinfo {author} {\bibfnamefont {S.}~\bibnamefont
  {Wehner}}, \bibinfo {author} {\bibfnamefont {D.}~\bibnamefont {Elkouss}}, \
  and\ \bibinfo {author} {\bibfnamefont {R.}~\bibnamefont {Hanson}},\ }\href
  {\doibase 10.1126/science.aam9288} {\bibfield  {journal} {\bibinfo  {journal}
  {Science}\ }\textbf {\bibinfo {volume} {362}},\ \bibinfo {pages} {eaam9288}
  (\bibinfo {year} {2018})}\BibitemShut {NoStop}%
\bibitem [{\citenamefont {Knill}\ \emph {et~al.}(2001)\citenamefont {Knill},
  \citenamefont {Laflamme},\ and\ \citenamefont {Milburn}}]{Knill2001Jan}%
  \BibitemOpen
  \bibfield  {author} {\bibinfo {author} {\bibfnamefont {E.}~\bibnamefont
  {Knill}}, \bibinfo {author} {\bibfnamefont {R.}~\bibnamefont {Laflamme}}, \
  and\ \bibinfo {author} {\bibfnamefont {G.~J.}\ \bibnamefont {Milburn}},\
  }\href {\doibase 10.1038/35051009} {\bibfield  {journal} {\bibinfo  {journal}
  {Nature}\ }\textbf {\bibinfo {volume} {409}},\ \bibinfo {pages} {46}
  (\bibinfo {year} {2001})}\BibitemShut {NoStop}%
\bibitem [{\citenamefont {Lvovsky}\ \emph {et~al.}(2009)\citenamefont
  {Lvovsky}, \citenamefont {Sanders},\ and\ \citenamefont
  {Tittel}}]{Lvovsky2009Dec}%
  \BibitemOpen
  \bibfield  {author} {\bibinfo {author} {\bibfnamefont {A.~I.}\ \bibnamefont
  {Lvovsky}}, \bibinfo {author} {\bibfnamefont {B.~C.}\ \bibnamefont
  {Sanders}}, \ and\ \bibinfo {author} {\bibfnamefont {W.}~\bibnamefont
  {Tittel}},\ }\href {\doibase 10.1038/nphoton.2009.231} {\bibfield  {journal}
  {\bibinfo  {journal} {Nat. Photonics}\ }\textbf {\bibinfo {volume} {3}},\
  \bibinfo {pages} {706} (\bibinfo {year} {2009})}\BibitemShut {NoStop}%
\bibitem [{\citenamefont {Tittel}\ \emph {et~al.}(2010)\citenamefont {Tittel},
  \citenamefont {Afzelius}, \citenamefont
  {Chaneli{\ifmmode\acute{e}\else\'{e}\fi}re}, \citenamefont {Cone},
  \citenamefont {Kr{\ifmmode\ddot{o}\else\"{o}\fi}ll}, \citenamefont
  {Moiseev},\ and\ \citenamefont {Sellars}}]{Tittel2010Feb}%
  \BibitemOpen
  \bibfield  {author} {\bibinfo {author} {\bibfnamefont {W.}~\bibnamefont
  {Tittel}}, \bibinfo {author} {\bibfnamefont {M.}~\bibnamefont {Afzelius}},
  \bibinfo {author} {\bibfnamefont {T.}~\bibnamefont
  {Chaneli{\ifmmode\acute{e}\else\'{e}\fi}re}}, \bibinfo {author}
  {\bibfnamefont {R.~L.}\ \bibnamefont {Cone}}, \bibinfo {author}
  {\bibfnamefont {S.}~\bibnamefont {Kr{\ifmmode\ddot{o}\else\"{o}\fi}ll}},
  \bibinfo {author} {\bibfnamefont {S.~A.}\ \bibnamefont {Moiseev}}, \ and\
  \bibinfo {author} {\bibfnamefont {M.}~\bibnamefont {Sellars}},\ }\href
  {\doibase 10.1002/lpor.200810056} {\bibfield  {journal} {\bibinfo  {journal}
  {Laser Photonics Rev.}\ }\textbf {\bibinfo {volume} {4}},\ \bibinfo {pages}
  {244} (\bibinfo {year} {2010})}\BibitemShut {NoStop}%
\bibitem [{\citenamefont {Zhong}\ \emph {et~al.}(2015)\citenamefont {Zhong},
  \citenamefont {Hedges}, \citenamefont {Ahlefeldt}, \citenamefont
  {Bartholomew}, \citenamefont {Beavan}, \citenamefont {Wittig}, \citenamefont
  {Longdell},\ and\ \citenamefont {Sellars}}]{Zhong2015Jan}%
  \BibitemOpen
  \bibfield  {author} {\bibinfo {author} {\bibfnamefont {M.}~\bibnamefont
  {Zhong}}, \bibinfo {author} {\bibfnamefont {M.~P.}\ \bibnamefont {Hedges}},
  \bibinfo {author} {\bibfnamefont {R.~L.}\ \bibnamefont {Ahlefeldt}}, \bibinfo
  {author} {\bibfnamefont {J.~G.}\ \bibnamefont {Bartholomew}}, \bibinfo
  {author} {\bibfnamefont {S.~E.}\ \bibnamefont {Beavan}}, \bibinfo {author}
  {\bibfnamefont {S.~M.}\ \bibnamefont {Wittig}}, \bibinfo {author}
  {\bibfnamefont {J.~J.}\ \bibnamefont {Longdell}}, \ and\ \bibinfo {author}
  {\bibfnamefont {M.~J.}\ \bibnamefont {Sellars}},\ }\href {\doibase
  10.1038/nature14025} {\bibfield  {journal} {\bibinfo  {journal} {Nature}\
  }\textbf {\bibinfo {volume} {517}},\ \bibinfo {pages} {177} (\bibinfo {year}
  {2015})}\BibitemShut {NoStop}%
\bibitem [{\citenamefont {Nilsson}\ \emph {et~al.}(2004)\citenamefont
  {Nilsson}, \citenamefont {Rippe}, \citenamefont {Kr\"oll}, \citenamefont
  {Klieber},\ and\ \citenamefont {Suter}}]{Nilsson2004}%
  \BibitemOpen
  \bibfield  {author} {\bibinfo {author} {\bibfnamefont {M.}~\bibnamefont
  {Nilsson}}, \bibinfo {author} {\bibfnamefont {L.}~\bibnamefont {Rippe}},
  \bibinfo {author} {\bibfnamefont {S.}~\bibnamefont {Kr\"oll}}, \bibinfo
  {author} {\bibfnamefont {R.}~\bibnamefont {Klieber}}, \ and\ \bibinfo
  {author} {\bibfnamefont {D.}~\bibnamefont {Suter}},\ }\href {\doibase
  10.1103/PhysRevB.70.214116} {\bibfield  {journal} {\bibinfo  {journal} {Phys.
  Rev. B}\ }\textbf {\bibinfo {volume} {70}},\ \bibinfo {pages} {214116}
  (\bibinfo {year} {2004})}\BibitemShut {NoStop}%
\bibitem [{\citenamefont {Sinclair}\ \emph {et~al.}(2014)\citenamefont
  {Sinclair}, \citenamefont {Saglamyurek}, \citenamefont {Mallahzadeh},
  \citenamefont {Slater}, \citenamefont {George}, \citenamefont {Ricken},
  \citenamefont {Hedges}, \citenamefont {Oblak}, \citenamefont {Simon},
  \citenamefont {Sohler},\ and\ \citenamefont {Tittel}}]{Sinclair2014Jul}%
  \BibitemOpen
  \bibfield  {author} {\bibinfo {author} {\bibfnamefont {N.}~\bibnamefont
  {Sinclair}}, \bibinfo {author} {\bibfnamefont {E.}~\bibnamefont
  {Saglamyurek}}, \bibinfo {author} {\bibfnamefont {H.}~\bibnamefont
  {Mallahzadeh}}, \bibinfo {author} {\bibfnamefont {J.~A.}\ \bibnamefont
  {Slater}}, \bibinfo {author} {\bibfnamefont {M.}~\bibnamefont {George}},
  \bibinfo {author} {\bibfnamefont {R.}~\bibnamefont {Ricken}}, \bibinfo
  {author} {\bibfnamefont {M.~P.}\ \bibnamefont {Hedges}}, \bibinfo {author}
  {\bibfnamefont {D.}~\bibnamefont {Oblak}}, \bibinfo {author} {\bibfnamefont
  {C.}~\bibnamefont {Simon}}, \bibinfo {author} {\bibfnamefont
  {W.}~\bibnamefont {Sohler}}, \ and\ \bibinfo {author} {\bibfnamefont
  {W.}~\bibnamefont {Tittel}},\ }\href {\doibase
  10.1103/PhysRevLett.113.053603} {\bibfield  {journal} {\bibinfo  {journal}
  {Phys. Rev. Lett.}\ }\textbf {\bibinfo {volume} {113}},\ \bibinfo {pages}
  {053603} (\bibinfo {year} {2014})}\BibitemShut {NoStop}%
\bibitem [{\citenamefont {Afzelius}\ \emph {et~al.}(2009)\citenamefont
  {Afzelius}, \citenamefont {Simon}, \citenamefont {de~Riedmatten},\ and\
  \citenamefont {Gisin}}]{Afzelius2009May}%
  \BibitemOpen
  \bibfield  {author} {\bibinfo {author} {\bibfnamefont {M.}~\bibnamefont
  {Afzelius}}, \bibinfo {author} {\bibfnamefont {C.}~\bibnamefont {Simon}},
  \bibinfo {author} {\bibfnamefont {H.}~\bibnamefont {de~Riedmatten}}, \ and\
  \bibinfo {author} {\bibfnamefont {N.}~\bibnamefont {Gisin}},\ }\href
  {\doibase 10.1103/PhysRevA.79.052329} {\bibfield  {journal} {\bibinfo
  {journal} {Phys. Rev. A}\ }\textbf {\bibinfo {volume} {79}},\ \bibinfo
  {pages} {052329} (\bibinfo {year} {2009})}\BibitemShut {NoStop}%
\bibitem [{\citenamefont {de~Riedmatten}\ \emph {et~al.}(2008)\citenamefont
  {de~Riedmatten}, \citenamefont {Afzelius}, \citenamefont {Staudt},
  \citenamefont {Simon},\ and\ \citenamefont {Gisin}}]{deRiedmatten2008Dec}%
  \BibitemOpen
  \bibfield  {author} {\bibinfo {author} {\bibfnamefont {H.}~\bibnamefont
  {de~Riedmatten}}, \bibinfo {author} {\bibfnamefont {M.}~\bibnamefont
  {Afzelius}}, \bibinfo {author} {\bibfnamefont {M.~U.}\ \bibnamefont
  {Staudt}}, \bibinfo {author} {\bibfnamefont {C.}~\bibnamefont {Simon}}, \
  and\ \bibinfo {author} {\bibfnamefont {N.}~\bibnamefont {Gisin}},\ }\href
  {\doibase 10.1038/nature07607} {\bibfield  {journal} {\bibinfo  {journal}
  {Nature}\ }\textbf {\bibinfo {volume} {456}},\ \bibinfo {pages} {773}
  (\bibinfo {year} {2008})}\BibitemShut {NoStop}%
\bibitem [{\citenamefont {Afzelius}\ \emph {et~al.}(2010)\citenamefont
  {Afzelius}, \citenamefont {Usmani}, \citenamefont {Amari}, \citenamefont
  {Lauritzen}, \citenamefont {Walther}, \citenamefont {Simon}, \citenamefont
  {Sangouard}, \citenamefont
  {Min{\ifmmode\acute{a}\else\'{a}\fi}{\ifmmode\check{r}\else\v{r}\fi}},
  \citenamefont {de~Riedmatten}, \citenamefont {Gisin},\ and\ \citenamefont
  {Kr{\ifmmode\ddot{o}\else\"{o}\fi}ll}}]{Afzelius2010Jan}%
  \BibitemOpen
  \bibfield  {author} {\bibinfo {author} {\bibfnamefont {M.}~\bibnamefont
  {Afzelius}}, \bibinfo {author} {\bibfnamefont {I.}~\bibnamefont {Usmani}},
  \bibinfo {author} {\bibfnamefont {A.}~\bibnamefont {Amari}}, \bibinfo
  {author} {\bibfnamefont {B.}~\bibnamefont {Lauritzen}}, \bibinfo {author}
  {\bibfnamefont {A.}~\bibnamefont {Walther}}, \bibinfo {author} {\bibfnamefont
  {C.}~\bibnamefont {Simon}}, \bibinfo {author} {\bibfnamefont
  {N.}~\bibnamefont {Sangouard}}, \bibinfo {author} {\bibfnamefont
  {J.}~\bibnamefont
  {Min{\ifmmode\acute{a}\else\'{a}\fi}{\ifmmode\check{r}\else\v{r}\fi}}},
  \bibinfo {author} {\bibfnamefont {H.}~\bibnamefont {de~Riedmatten}}, \bibinfo
  {author} {\bibfnamefont {N.}~\bibnamefont {Gisin}}, \ and\ \bibinfo {author}
  {\bibfnamefont {S.}~\bibnamefont {Kr{\ifmmode\ddot{o}\else\"{o}\fi}ll}},\
  }\href {\doibase 10.1103/PhysRevLett.104.040503} {\bibfield  {journal}
  {\bibinfo  {journal} {Phys. Rev. Lett.}\ }\textbf {\bibinfo {volume} {104}},\
  \bibinfo {pages} {040503} (\bibinfo {year} {2010})}\BibitemShut {NoStop}%
\bibitem [{\citenamefont {Afzelius}\ and\ \citenamefont
  {Simon}(2010)}]{Afzelius2010Aug}%
  \BibitemOpen
  \bibfield  {author} {\bibinfo {author} {\bibfnamefont {M.}~\bibnamefont
  {Afzelius}}\ and\ \bibinfo {author} {\bibfnamefont {C.}~\bibnamefont
  {Simon}},\ }\href {\doibase 10.1103/PhysRevA.82.022310} {\bibfield  {journal}
  {\bibinfo  {journal} {Phys. Rev. A}\ }\textbf {\bibinfo {volume} {82}},\
  \bibinfo {pages} {022310} (\bibinfo {year} {2010})}\BibitemShut {NoStop}%
\bibitem [{\citenamefont {Sabooni}\ \emph {et~al.}(2013)\citenamefont
  {Sabooni}, \citenamefont {Li}, \citenamefont
  {Kr{\ifmmode\ddot{o}\else\"{o}\fi}ll},\ and\ \citenamefont
  {Rippe}}]{Sabooni2013Mar}%
  \BibitemOpen
  \bibfield  {author} {\bibinfo {author} {\bibfnamefont {M.}~\bibnamefont
  {Sabooni}}, \bibinfo {author} {\bibfnamefont {Q.}~\bibnamefont {Li}},
  \bibinfo {author} {\bibfnamefont {S.}~\bibnamefont
  {Kr{\ifmmode\ddot{o}\else\"{o}\fi}ll}}, \ and\ \bibinfo {author}
  {\bibfnamefont {L.}~\bibnamefont {Rippe}},\ }\href {\doibase
  10.1103/PhysRevLett.110.133604} {\bibfield  {journal} {\bibinfo  {journal}
  {Phys. Rev. Lett.}\ }\textbf {\bibinfo {volume} {110}},\ \bibinfo {pages}
  {133604} (\bibinfo {year} {2013})}\BibitemShut {NoStop}%
\bibitem [{\citenamefont {Jobez}\ \emph {et~al.}(2014)\citenamefont {Jobez},
  \citenamefont {Usmani}, \citenamefont {Timoney}, \citenamefont {Laplane},
  \citenamefont {Gisin},\ and\ \citenamefont {Afzelius}}]{Jobez2014Aug}%
  \BibitemOpen
  \bibfield  {author} {\bibinfo {author} {\bibfnamefont {P.}~\bibnamefont
  {Jobez}}, \bibinfo {author} {\bibfnamefont {I.}~\bibnamefont {Usmani}},
  \bibinfo {author} {\bibfnamefont {N.}~\bibnamefont {Timoney}}, \bibinfo
  {author} {\bibfnamefont {C.}~\bibnamefont {Laplane}}, \bibinfo {author}
  {\bibfnamefont {N.}~\bibnamefont {Gisin}}, \ and\ \bibinfo {author}
  {\bibfnamefont {M.}~\bibnamefont {Afzelius}},\ }\href {\doibase
  10.1088/1367-2630/16/8/083005} {\bibfield  {journal} {\bibinfo  {journal}
  {New J. Phys.}\ }\textbf {\bibinfo {volume} {16}},\ \bibinfo {pages} {083005}
  (\bibinfo {year} {2014})}\BibitemShut {NoStop}%
\bibitem [{\citenamefont
  {G{\ifmmode\ddot{u}\else\"{u}\fi}ndo{\ifmmode\breve{g}\else\u{g}\fi}an}\
  \emph {et~al.}(2015)\citenamefont
  {G{\ifmmode\ddot{u}\else\"{u}\fi}ndo{\ifmmode\breve{g}\else\u{g}\fi}an},
  \citenamefont {Ledingham}, \citenamefont {Kutluer}, \citenamefont {Mazzera},\
  and\ \citenamefont {de~Riedmatten}}]{Gundogan2015Jun}%
  \BibitemOpen
  \bibfield  {author} {\bibinfo {author} {\bibfnamefont {M.}~\bibnamefont
  {G{\ifmmode\ddot{u}\else\"{u}\fi}ndo{\ifmmode\breve{g}\else\u{g}\fi}an}},
  \bibinfo {author} {\bibfnamefont {P.~M.}\ \bibnamefont {Ledingham}}, \bibinfo
  {author} {\bibfnamefont {K.}~\bibnamefont {Kutluer}}, \bibinfo {author}
  {\bibfnamefont {M.}~\bibnamefont {Mazzera}}, \ and\ \bibinfo {author}
  {\bibfnamefont {H.}~\bibnamefont {de~Riedmatten}},\ }\href {\doibase
  10.1103/PhysRevLett.114.230501} {\bibfield  {journal} {\bibinfo  {journal}
  {Phys. Rev. Lett.}\ }\textbf {\bibinfo {volume} {114}},\ \bibinfo {pages}
  {230501} (\bibinfo {year} {2015})}\BibitemShut {NoStop}%
\bibitem [{\citenamefont {Timoney}\ \emph {et~al.}(2013)\citenamefont
  {Timoney}, \citenamefont {Usmani}, \citenamefont {Jobez}, \citenamefont
  {Afzelius},\ and\ \citenamefont {Gisin}}]{Timoney2013Aug}%
  \BibitemOpen
  \bibfield  {author} {\bibinfo {author} {\bibfnamefont {N.}~\bibnamefont
  {Timoney}}, \bibinfo {author} {\bibfnamefont {I.}~\bibnamefont {Usmani}},
  \bibinfo {author} {\bibfnamefont {P.}~\bibnamefont {Jobez}}, \bibinfo
  {author} {\bibfnamefont {M.}~\bibnamefont {Afzelius}}, \ and\ \bibinfo
  {author} {\bibfnamefont {N.}~\bibnamefont {Gisin}},\ }\href {\doibase
  10.1103/PhysRevA.88.022324} {\bibfield  {journal} {\bibinfo  {journal} {Phys.
  Rev. A}\ }\textbf {\bibinfo {volume} {88}},\ \bibinfo {pages} {022324}
  (\bibinfo {year} {2013})}\BibitemShut {NoStop}%
\bibitem [{\citenamefont {Jobez}\ \emph {et~al.}(2015)\citenamefont {Jobez},
  \citenamefont {Laplane}, \citenamefont {Timoney}, \citenamefont {Gisin},
  \citenamefont {Ferrier}, \citenamefont {Goldner},\ and\ \citenamefont
  {Afzelius}}]{Jobez2015Jun}%
  \BibitemOpen
  \bibfield  {author} {\bibinfo {author} {\bibfnamefont {P.}~\bibnamefont
  {Jobez}}, \bibinfo {author} {\bibfnamefont {C.}~\bibnamefont {Laplane}},
  \bibinfo {author} {\bibfnamefont {N.}~\bibnamefont {Timoney}}, \bibinfo
  {author} {\bibfnamefont {N.}~\bibnamefont {Gisin}}, \bibinfo {author}
  {\bibfnamefont {A.}~\bibnamefont {Ferrier}}, \bibinfo {author} {\bibfnamefont
  {P.}~\bibnamefont {Goldner}}, \ and\ \bibinfo {author} {\bibfnamefont
  {M.}~\bibnamefont {Afzelius}},\ }\href {\doibase
  10.1103/PhysRevLett.114.230502} {\bibfield  {journal} {\bibinfo  {journal}
  {Phys. Rev. Lett.}\ }\textbf {\bibinfo {volume} {114}},\ \bibinfo {pages}
  {230502} (\bibinfo {year} {2015})}\BibitemShut {NoStop}%
\bibitem [{\citenamefont {Bonarota}\ \emph {et~al.}(2014)\citenamefont
  {Bonarota}, \citenamefont {Dajczgewand}, \citenamefont {Louchet-Chauvet},
  \citenamefont {Le~Gou{\ifmmode\ddot{e}\else\"{e}\fi}t},\ and\ \citenamefont
  {Chaneli{\ifmmode\grave{e}\else\`{e}\fi}re}}]{Bonarota2014Aug}%
  \BibitemOpen
  \bibfield  {author} {\bibinfo {author} {\bibfnamefont {M.}~\bibnamefont
  {Bonarota}}, \bibinfo {author} {\bibfnamefont {J.}~\bibnamefont
  {Dajczgewand}}, \bibinfo {author} {\bibfnamefont {A.}~\bibnamefont
  {Louchet-Chauvet}}, \bibinfo {author} {\bibfnamefont {J.-L.}\ \bibnamefont
  {Le~Gou{\ifmmode\ddot{e}\else\"{e}\fi}t}}, \ and\ \bibinfo {author}
  {\bibfnamefont {T.}~\bibnamefont
  {Chaneli{\ifmmode\grave{e}\else\`{e}\fi}re}},\ }\href {\doibase
  10.1088/1054-660x/24/9/094003} {\bibfield  {journal} {\bibinfo  {journal}
  {Laser Phys.}\ }\textbf {\bibinfo {volume} {24}},\ \bibinfo {pages} {094003}
  (\bibinfo {year} {2014})}\BibitemShut {NoStop}%
\bibitem [{\citenamefont {Horvath}\ \emph {et~al.}(2021)\citenamefont
  {Horvath}, \citenamefont {Alqedra}, \citenamefont {Kinos}, \citenamefont
  {Walther}, \citenamefont {Dahlstr{\ifmmode\ddot{o}\else\"{o}\fi}m},
  \citenamefont {Kr{\ifmmode\ddot{o}\else\"{o}\fi}ll},\ and\ \citenamefont
  {Rippe}}]{Horvath2021May}%
  \BibitemOpen
  \bibfield  {author} {\bibinfo {author} {\bibfnamefont {S.~P.}\ \bibnamefont
  {Horvath}}, \bibinfo {author} {\bibfnamefont {M.~K.}\ \bibnamefont
  {Alqedra}}, \bibinfo {author} {\bibfnamefont {A.}~\bibnamefont {Kinos}},
  \bibinfo {author} {\bibfnamefont {A.}~\bibnamefont {Walther}}, \bibinfo
  {author} {\bibfnamefont {J.~M.}\ \bibnamefont
  {Dahlstr{\ifmmode\ddot{o}\else\"{o}\fi}m}}, \bibinfo {author} {\bibfnamefont
  {S.}~\bibnamefont {Kr{\ifmmode\ddot{o}\else\"{o}\fi}ll}}, \ and\ \bibinfo
  {author} {\bibfnamefont {L.}~\bibnamefont {Rippe}},\ }\href {\doibase
  10.1103/PhysRevResearch.3.023099} {\bibfield  {journal} {\bibinfo  {journal}
  {Phys. Rev. Res.}\ }\textbf {\bibinfo {volume} {3}},\ \bibinfo {pages}
  {023099} (\bibinfo {year} {2021})}\BibitemShut {NoStop}%
\bibitem [{\citenamefont {Meixner}\ \emph {et~al.}(1992)\citenamefont
  {Meixner}, \citenamefont {Jefferson},\ and\ \citenamefont
  {Macfarlane}}]{Meixner1992Sep}%
  \BibitemOpen
  \bibfield  {author} {\bibinfo {author} {\bibfnamefont {A.~J.}\ \bibnamefont
  {Meixner}}, \bibinfo {author} {\bibfnamefont {C.~M.}\ \bibnamefont
  {Jefferson}}, \ and\ \bibinfo {author} {\bibfnamefont {R.~M.}\ \bibnamefont
  {Macfarlane}},\ }\href {\doibase 10.1103/PhysRevB.46.5912} {\bibfield
  {journal} {\bibinfo  {journal} {Phys. Rev. B}\ }\textbf {\bibinfo {volume}
  {46}},\ \bibinfo {pages} {5912} (\bibinfo {year} {1992})}\BibitemShut
  {NoStop}%
\bibitem [{\citenamefont {Wang}\ and\ \citenamefont
  {Meltzer}(1992)}]{Wang1992May}%
  \BibitemOpen
  \bibfield  {author} {\bibinfo {author} {\bibfnamefont {Y.~P.}\ \bibnamefont
  {Wang}}\ and\ \bibinfo {author} {\bibfnamefont {R.~S.}\ \bibnamefont
  {Meltzer}},\ }\href {\doibase 10.1103/PhysRevB.45.10119} {\bibfield
  {journal} {\bibinfo  {journal} {Phys. Rev. B}\ }\textbf {\bibinfo {volume}
  {45}},\ \bibinfo {pages} {10119} (\bibinfo {year} {1992})}\BibitemShut
  {NoStop}%
\bibitem [{\citenamefont {Graf}\ \emph
  {et~al.}(1997{\natexlab{a}})\citenamefont {Graf}, \citenamefont {Renn},
  \citenamefont {Wild},\ and\ \citenamefont {Mitsunaga}}]{Graf1997}%
  \BibitemOpen
  \bibfield  {author} {\bibinfo {author} {\bibfnamefont {F.~R.}\ \bibnamefont
  {Graf}}, \bibinfo {author} {\bibfnamefont {A.}~\bibnamefont {Renn}}, \bibinfo
  {author} {\bibfnamefont {U.~P.}\ \bibnamefont {Wild}}, \ and\ \bibinfo
  {author} {\bibfnamefont {M.}~\bibnamefont {Mitsunaga}},\ }\href {\doibase
  10.1103/PhysRevB.55.11225} {\bibfield  {journal} {\bibinfo  {journal} {Phys.
  Rev. B}\ }\textbf {\bibinfo {volume} {55}},\ \bibinfo {pages} {11225}
  (\bibinfo {year} {1997}{\natexlab{a}})}\BibitemShut {NoStop}%
\bibitem [{\citenamefont {Chaneli{\ifmmode\grave{e}\else\`{e}\fi}re}\ \emph
  {et~al.}(2008)\citenamefont {Chaneli{\ifmmode\grave{e}\else\`{e}\fi}re},
  \citenamefont {Ruggiero}, \citenamefont
  {Le~Gou{\ifmmode\ddot{e}\else\"{e}\fi}t}, \citenamefont {Tittel},
  \citenamefont {Mun}, \citenamefont {Jouini}, \citenamefont {Yoshikawa},
  \citenamefont {Boulon}, \citenamefont {Du}, \citenamefont {Goldner},
  \citenamefont {Beaudoux}, \citenamefont {Vincent}, \citenamefont
  {Antic-Fidancev},\ and\ \citenamefont
  {Guillot-No{\ifmmode\ddot{e}\else\"{e}\fi}l}}]{Chaneliere2008Jun}%
  \BibitemOpen
  \bibfield  {author} {\bibinfo {author} {\bibfnamefont {T.}~\bibnamefont
  {Chaneli{\ifmmode\grave{e}\else\`{e}\fi}re}}, \bibinfo {author}
  {\bibfnamefont {J.}~\bibnamefont {Ruggiero}}, \bibinfo {author}
  {\bibfnamefont {J.-L.}\ \bibnamefont
  {Le~Gou{\ifmmode\ddot{e}\else\"{e}\fi}t}}, \bibinfo {author} {\bibfnamefont
  {W.}~\bibnamefont {Tittel}}, \bibinfo {author} {\bibfnamefont {J.-H.}\
  \bibnamefont {Mun}}, \bibinfo {author} {\bibfnamefont {A.}~\bibnamefont
  {Jouini}}, \bibinfo {author} {\bibfnamefont {A.}~\bibnamefont {Yoshikawa}},
  \bibinfo {author} {\bibfnamefont {G.}~\bibnamefont {Boulon}}, \bibinfo
  {author} {\bibfnamefont {Y.~L.}\ \bibnamefont {Du}}, \bibinfo {author}
  {\bibfnamefont {{\relax Ph}.}~\bibnamefont {Goldner}}, \bibinfo {author}
  {\bibfnamefont {F.}~\bibnamefont {Beaudoux}}, \bibinfo {author}
  {\bibfnamefont {J.}~\bibnamefont {Vincent}}, \bibinfo {author} {\bibfnamefont
  {E.}~\bibnamefont {Antic-Fidancev}}, \ and\ \bibinfo {author} {\bibfnamefont
  {O.}~\bibnamefont {Guillot-No{\ifmmode\ddot{e}\else\"{e}\fi}l}},\ }\href
  {\doibase 10.1103/PhysRevB.77.245127} {\bibfield  {journal} {\bibinfo
  {journal} {Phys. Rev. B}\ }\textbf {\bibinfo {volume} {77}},\ \bibinfo
  {pages} {245127} (\bibinfo {year} {2008})}\BibitemShut {NoStop}%
\bibitem [{\citenamefont {Arcangeli}\ \emph {et~al.}(2016)\citenamefont
  {Arcangeli}, \citenamefont {Ferrier},\ and\ \citenamefont
  {Goldner}}]{Arcangeli2016Jun}%
  \BibitemOpen
  \bibfield  {author} {\bibinfo {author} {\bibfnamefont {A.}~\bibnamefont
  {Arcangeli}}, \bibinfo {author} {\bibfnamefont {A.}~\bibnamefont {Ferrier}},
  \ and\ \bibinfo {author} {\bibfnamefont {{\relax Ph}.}~\bibnamefont
  {Goldner}},\ }\href {\doibase 10.1103/PhysRevA.93.062303} {\bibfield
  {journal} {\bibinfo  {journal} {Phys. Rev. A}\ }\textbf {\bibinfo {volume}
  {93}},\ \bibinfo {pages} {062303} (\bibinfo {year} {2016})}\BibitemShut
  {NoStop}%
\bibitem [{\citenamefont {Nilsson}\ and\ \citenamefont
  {Kr{\ifmmode\ddot{o}\else\"{o}\fi}ll}(2005)}]{Nilsson2005Mar}%
  \BibitemOpen
  \bibfield  {author} {\bibinfo {author} {\bibfnamefont {M.}~\bibnamefont
  {Nilsson}}\ and\ \bibinfo {author} {\bibfnamefont {S.}~\bibnamefont
  {Kr{\ifmmode\ddot{o}\else\"{o}\fi}ll}},\ }\href {\doibase
  10.1016/j.optcom.2004.11.077} {\bibfield  {journal} {\bibinfo  {journal}
  {Opt. Commun.}\ }\textbf {\bibinfo {volume} {247}},\ \bibinfo {pages} {393}
  (\bibinfo {year} {2005})}\BibitemShut {NoStop}%
\bibitem [{\citenamefont {Alexander}\ \emph {et~al.}(2006)\citenamefont
  {Alexander}, \citenamefont {Longdell}, \citenamefont {Sellars},\ and\
  \citenamefont {Manson}}]{Alexander2006Feb}%
  \BibitemOpen
  \bibfield  {author} {\bibinfo {author} {\bibfnamefont {A.~L.}\ \bibnamefont
  {Alexander}}, \bibinfo {author} {\bibfnamefont {J.~J.}\ \bibnamefont
  {Longdell}}, \bibinfo {author} {\bibfnamefont {M.~J.}\ \bibnamefont
  {Sellars}}, \ and\ \bibinfo {author} {\bibfnamefont {N.~B.}\ \bibnamefont
  {Manson}},\ }\href {\doibase 10.1103/PhysRevLett.96.043602} {\bibfield
  {journal} {\bibinfo  {journal} {Phys. Rev. Lett.}\ }\textbf {\bibinfo
  {volume} {96}},\ \bibinfo {pages} {043602} (\bibinfo {year}
  {2006})}\BibitemShut {NoStop}%
\bibitem [{\citenamefont {Kraus}\ \emph {et~al.}(2006)\citenamefont {Kraus},
  \citenamefont {Tittel}, \citenamefont {Gisin}, \citenamefont {Nilsson},
  \citenamefont {Kr{\ifmmode\ddot{o}\else\"{o}\fi}ll},\ and\ \citenamefont
  {Cirac}}]{Kraus2006Feb}%
  \BibitemOpen
  \bibfield  {author} {\bibinfo {author} {\bibfnamefont {B.}~\bibnamefont
  {Kraus}}, \bibinfo {author} {\bibfnamefont {W.}~\bibnamefont {Tittel}},
  \bibinfo {author} {\bibfnamefont {N.}~\bibnamefont {Gisin}}, \bibinfo
  {author} {\bibfnamefont {M.}~\bibnamefont {Nilsson}}, \bibinfo {author}
  {\bibfnamefont {S.}~\bibnamefont {Kr{\ifmmode\ddot{o}\else\"{o}\fi}ll}}, \
  and\ \bibinfo {author} {\bibfnamefont {J.~I.}\ \bibnamefont {Cirac}},\ }\href
  {\doibase 10.1103/PhysRevA.73.020302} {\bibfield  {journal} {\bibinfo
  {journal} {Phys. Rev. A}\ }\textbf {\bibinfo {volume} {73}},\ \bibinfo
  {pages} {020302} (\bibinfo {year} {2006})}\BibitemShut {NoStop}%
\bibitem [{\citenamefont {Lauritzen}\ \emph {et~al.}(2011)\citenamefont
  {Lauritzen}, \citenamefont
  {Min{\ifmmode\acute{a}\else\'{a}\fi}{\ifmmode\check{r}\else\v{r}\fi}},
  \citenamefont {de~Riedmatten}, \citenamefont {Afzelius},\ and\ \citenamefont
  {Gisin}}]{Lauritzen2011Jan}%
  \BibitemOpen
  \bibfield  {author} {\bibinfo {author} {\bibfnamefont {B.}~\bibnamefont
  {Lauritzen}}, \bibinfo {author} {\bibfnamefont {J.}~\bibnamefont
  {Min{\ifmmode\acute{a}\else\'{a}\fi}{\ifmmode\check{r}\else\v{r}\fi}}},
  \bibinfo {author} {\bibfnamefont {H.}~\bibnamefont {de~Riedmatten}}, \bibinfo
  {author} {\bibfnamefont {M.}~\bibnamefont {Afzelius}}, \ and\ \bibinfo
  {author} {\bibfnamefont {N.}~\bibnamefont {Gisin}},\ }\href {\doibase
  10.1103/PhysRevA.83.012318} {\bibfield  {journal} {\bibinfo  {journal} {Phys.
  Rev. A}\ }\textbf {\bibinfo {volume} {83}},\ \bibinfo {pages} {012318}
  (\bibinfo {year} {2011})}\BibitemShut {NoStop}%
\bibitem [{Note1()}]{Note1}%
  \BibitemOpen
  \bibinfo {note} {What happens in Vegas, stays in Vegas!}\BibitemShut {Stop}%
\bibitem [{\citenamefont {Graf}\ \emph
  {et~al.}(1997{\natexlab{b}})\citenamefont {Graf}, \citenamefont {Renn},
  \citenamefont {Wild},\ and\ \citenamefont {Mitsunaga}}]{Graf1997May}%
  \BibitemOpen
  \bibfield  {author} {\bibinfo {author} {\bibfnamefont {F.~R.}\ \bibnamefont
  {Graf}}, \bibinfo {author} {\bibfnamefont {A.}~\bibnamefont {Renn}}, \bibinfo
  {author} {\bibfnamefont {U.~P.}\ \bibnamefont {Wild}}, \ and\ \bibinfo
  {author} {\bibfnamefont {M.}~\bibnamefont {Mitsunaga}},\ }\href {\doibase
  10.1103/PhysRevB.55.11225} {\bibfield  {journal} {\bibinfo  {journal} {Phys.
  Rev. B}\ }\textbf {\bibinfo {volume} {55}},\ \bibinfo {pages} {11225}
  (\bibinfo {year} {1997}{\natexlab{b}})}\BibitemShut {NoStop}%
\bibitem [{\citenamefont {Amari}\ \emph {et~al.}(2010)\citenamefont {Amari},
  \citenamefont {Walther}, \citenamefont {Sabooni}, \citenamefont {Huang},
  \citenamefont {Kr{\ifmmode\ddot{o}\else\"{o}\fi}ll}, \citenamefont
  {Afzelius}, \citenamefont {Usmani}, \citenamefont {Lauritzen}, \citenamefont
  {Sangouard}, \citenamefont {de~Riedmatten},\ and\ \citenamefont
  {Gisin}}]{Amari2010Sep}%
  \BibitemOpen
  \bibfield  {author} {\bibinfo {author} {\bibfnamefont {A.}~\bibnamefont
  {Amari}}, \bibinfo {author} {\bibfnamefont {A.}~\bibnamefont {Walther}},
  \bibinfo {author} {\bibfnamefont {M.}~\bibnamefont {Sabooni}}, \bibinfo
  {author} {\bibfnamefont {M.}~\bibnamefont {Huang}}, \bibinfo {author}
  {\bibfnamefont {S.}~\bibnamefont {Kr{\ifmmode\ddot{o}\else\"{o}\fi}ll}},
  \bibinfo {author} {\bibfnamefont {M.}~\bibnamefont {Afzelius}}, \bibinfo
  {author} {\bibfnamefont {I.}~\bibnamefont {Usmani}}, \bibinfo {author}
  {\bibfnamefont {B.}~\bibnamefont {Lauritzen}}, \bibinfo {author}
  {\bibfnamefont {N.}~\bibnamefont {Sangouard}}, \bibinfo {author}
  {\bibfnamefont {H.}~\bibnamefont {de~Riedmatten}}, \ and\ \bibinfo {author}
  {\bibfnamefont {N.}~\bibnamefont {Gisin}},\ }\href {\doibase
  10.1016/j.jlumin.2010.01.012} {\bibfield  {journal} {\bibinfo  {journal} {J.
  Lumin.}\ }\textbf {\bibinfo {volume} {130}},\ \bibinfo {pages} {1579}
  (\bibinfo {year} {2010})}\BibitemShut {NoStop}%
\bibitem [{\citenamefont {Rippe}\ \emph {et~al.}(2005)\citenamefont {Rippe},
  \citenamefont {Nilsson}, \citenamefont {Kr{\ifmmode\ddot{o}\else\"{o}\fi}ll},
  \citenamefont {Klieber},\ and\ \citenamefont {Suter}}]{Rippe2005Jun}%
  \BibitemOpen
  \bibfield  {author} {\bibinfo {author} {\bibfnamefont {L.}~\bibnamefont
  {Rippe}}, \bibinfo {author} {\bibfnamefont {M.}~\bibnamefont {Nilsson}},
  \bibinfo {author} {\bibfnamefont {S.}~\bibnamefont
  {Kr{\ifmmode\ddot{o}\else\"{o}\fi}ll}}, \bibinfo {author} {\bibfnamefont
  {R.}~\bibnamefont {Klieber}}, \ and\ \bibinfo {author} {\bibfnamefont
  {D.}~\bibnamefont {Suter}},\ }\href {\doibase 10.1103/PhysRevA.71.062328}
  {\bibfield  {journal} {\bibinfo  {journal} {Phys. Rev. A}\ }\textbf {\bibinfo
  {volume} {71}},\ \bibinfo {pages} {062328} (\bibinfo {year}
  {2005})}\BibitemShut {NoStop}%
\bibitem [{\citenamefont {Roos}\ and\ \citenamefont
  {M{\o}lmer}(2004)}]{Roos2004Feb}%
  \BibitemOpen
  \bibfield  {author} {\bibinfo {author} {\bibfnamefont {I.}~\bibnamefont
  {Roos}}\ and\ \bibinfo {author} {\bibfnamefont {K.}~\bibnamefont
  {M{\o}lmer}},\ }\href {\doibase 10.1103/PhysRevA.69.022321} {\bibfield
  {journal} {\bibinfo  {journal} {Phys. Rev. A}\ }\textbf {\bibinfo {volume}
  {69}},\ \bibinfo {pages} {022321} (\bibinfo {year} {2004})}\BibitemShut
  {NoStop}%
\bibitem [{\citenamefont {Timoney}\ \emph {et~al.}(2012)\citenamefont
  {Timoney}, \citenamefont {Lauritzen}, \citenamefont {Usmani}, \citenamefont
  {Afzelius},\ and\ \citenamefont {Gisin}}]{Timoney2012Jun}%
  \BibitemOpen
  \bibfield  {author} {\bibinfo {author} {\bibfnamefont {N.}~\bibnamefont
  {Timoney}}, \bibinfo {author} {\bibfnamefont {B.}~\bibnamefont {Lauritzen}},
  \bibinfo {author} {\bibfnamefont {I.}~\bibnamefont {Usmani}}, \bibinfo
  {author} {\bibfnamefont {M.}~\bibnamefont {Afzelius}}, \ and\ \bibinfo
  {author} {\bibfnamefont {N.}~\bibnamefont {Gisin}},\ }\href {\doibase
  10.1088/0953-4075/45/12/124001} {\bibfield  {journal} {\bibinfo  {journal}
  {J. Phys. B: At. Mol. Opt. Phys.}\ }\textbf {\bibinfo {volume} {45}},\
  \bibinfo {pages} {124001} (\bibinfo {year} {2012})}\BibitemShut {NoStop}%
\bibitem [{\citenamefont
  {Min{\ifmmode\acute{a}\else\'{a}\fi}{\ifmmode\check{r}\else\v{r}\fi}}\ \emph
  {et~al.}(2009)\citenamefont
  {Min{\ifmmode\acute{a}\else\'{a}\fi}{\ifmmode\check{r}\else\v{r}\fi}},
  \citenamefont {Lauritzen}, \citenamefont {de~Riedmatten}, \citenamefont
  {Afzelius}, \citenamefont {Simon},\ and\ \citenamefont
  {Gisin}}]{Minar2009Nov}%
  \BibitemOpen
  \bibfield  {author} {\bibinfo {author} {\bibfnamefont {J.}~\bibnamefont
  {Min{\ifmmode\acute{a}\else\'{a}\fi}{\ifmmode\check{r}\else\v{r}\fi}}},
  \bibinfo {author} {\bibfnamefont {B.}~\bibnamefont {Lauritzen}}, \bibinfo
  {author} {\bibfnamefont {H.}~\bibnamefont {de~Riedmatten}}, \bibinfo {author}
  {\bibfnamefont {M.}~\bibnamefont {Afzelius}}, \bibinfo {author}
  {\bibfnamefont {C.}~\bibnamefont {Simon}}, \ and\ \bibinfo {author}
  {\bibfnamefont {N.}~\bibnamefont {Gisin}},\ }\href {\doibase
  10.1088/1367-2630/11/11/113019} {\bibfield  {journal} {\bibinfo  {journal}
  {New J. Phys.}\ }\textbf {\bibinfo {volume} {11}},\ \bibinfo {pages} {113019}
  (\bibinfo {year} {2009})}\BibitemShut {NoStop}%
\bibitem [{\citenamefont {Horvath}\ \emph {et~al.}(2022)\citenamefont
  {Horvath}, \citenamefont {Shi}, \citenamefont {Gustavsson}, \citenamefont
  {Walther}, \citenamefont {Kinos}, \citenamefont
  {Kr{\ifmmode\ddot{o}\else\"{o}\fi}ll},\ and\ \citenamefont
  {Rippe}}]{Horvath2022Mar}%
  \BibitemOpen
  \bibfield  {author} {\bibinfo {author} {\bibfnamefont {S.~P.}\ \bibnamefont
  {Horvath}}, \bibinfo {author} {\bibfnamefont {C.}~\bibnamefont {Shi}},
  \bibinfo {author} {\bibfnamefont {D.}~\bibnamefont {Gustavsson}}, \bibinfo
  {author} {\bibfnamefont {A.}~\bibnamefont {Walther}}, \bibinfo {author}
  {\bibfnamefont {A.}~\bibnamefont {Kinos}}, \bibinfo {author} {\bibfnamefont
  {S.}~\bibnamefont {Kr{\ifmmode\ddot{o}\else\"{o}\fi}ll}}, \ and\ \bibinfo
  {author} {\bibfnamefont {L.}~\bibnamefont {Rippe}},\ }\href {\doibase
  10.1088/1367-2630/ac5932} {\bibfield  {journal} {\bibinfo  {journal} {New J.
  Phys.}\ }\textbf {\bibinfo {volume} {24}},\ \bibinfo {pages} {033034}
  (\bibinfo {year} {2022})}\BibitemShut {NoStop}%
\bibitem [{\citenamefont {Equall}\ \emph {et~al.}(1994)\citenamefont {Equall},
  \citenamefont {Sun}, \citenamefont {Cone},\ and\ \citenamefont
  {Macfarlane}}]{Equall1994Apr}%
  \BibitemOpen
  \bibfield  {author} {\bibinfo {author} {\bibfnamefont {R.~W.}\ \bibnamefont
  {Equall}}, \bibinfo {author} {\bibfnamefont {Y.}~\bibnamefont {Sun}},
  \bibinfo {author} {\bibfnamefont {R.~L.}\ \bibnamefont {Cone}}, \ and\
  \bibinfo {author} {\bibfnamefont {R.~M.}\ \bibnamefont {Macfarlane}},\ }\href
  {\doibase 10.1103/PhysRevLett.72.2179} {\bibfield  {journal} {\bibinfo
  {journal} {Phys. Rev. Lett.}\ }\textbf {\bibinfo {volume} {72}},\ \bibinfo
  {pages} {2179} (\bibinfo {year} {1994})}\BibitemShut {NoStop}%
\bibitem [{\citenamefont {Hilborn}(1998)}]{Hilborn1998Jul}%
  \BibitemOpen
  \bibfield  {author} {\bibinfo {author} {\bibfnamefont {R.~C.}\ \bibnamefont
  {Hilborn}},\ }\href {\doibase 10.1119/1.12937} {\bibfield  {journal}
  {\bibinfo  {journal} {Am. J. Phys.}\ }\textbf {\bibinfo {volume} {50}},\
  \bibinfo {pages} {982} (\bibinfo {year} {1998})}\BibitemShut {NoStop}%
\bibitem [{\citenamefont {Graf}\ \emph {et~al.}(1998)\citenamefont {Graf},
  \citenamefont {Renn}, \citenamefont {Zumofen},\ and\ \citenamefont
  {Wild}}]{Graf1998Sep}%
  \BibitemOpen
  \bibfield  {author} {\bibinfo {author} {\bibfnamefont {F.~R.}\ \bibnamefont
  {Graf}}, \bibinfo {author} {\bibfnamefont {A.}~\bibnamefont {Renn}}, \bibinfo
  {author} {\bibfnamefont {G.}~\bibnamefont {Zumofen}}, \ and\ \bibinfo
  {author} {\bibfnamefont {U.~P.}\ \bibnamefont {Wild}},\ }\href {\doibase
  10.1103/PhysRevB.58.5462} {\bibfield  {journal} {\bibinfo  {journal} {Phys.
  Rev. B}\ }\textbf {\bibinfo {volume} {58}},\ \bibinfo {pages} {5462}
  (\bibinfo {year} {1998})}\BibitemShut {NoStop}%
\bibitem [{\citenamefont {K{\ifmmode\ddot{o}\else\"{o}\fi}nz}\ \emph
  {et~al.}(2003)\citenamefont {K{\ifmmode\ddot{o}\else\"{o}\fi}nz},
  \citenamefont {Sun}, \citenamefont {Thiel}, \citenamefont {Cone},
  \citenamefont {Equall}, \citenamefont {Hutcheson},\ and\ \citenamefont
  {Macfarlane}}]{Konz2003Aug}%
  \BibitemOpen
  \bibfield  {author} {\bibinfo {author} {\bibfnamefont {F.}~\bibnamefont
  {K{\ifmmode\ddot{o}\else\"{o}\fi}nz}}, \bibinfo {author} {\bibfnamefont
  {Y.}~\bibnamefont {Sun}}, \bibinfo {author} {\bibfnamefont {C.~W.}\
  \bibnamefont {Thiel}}, \bibinfo {author} {\bibfnamefont {R.~L.}\ \bibnamefont
  {Cone}}, \bibinfo {author} {\bibfnamefont {R.~W.}\ \bibnamefont {Equall}},
  \bibinfo {author} {\bibfnamefont {R.~L.}\ \bibnamefont {Hutcheson}}, \ and\
  \bibinfo {author} {\bibfnamefont {R.~M.}\ \bibnamefont {Macfarlane}},\ }\href
  {\doibase 10.1103/PhysRevB.68.085109} {\bibfield  {journal} {\bibinfo
  {journal} {Phys. Rev. B}\ }\textbf {\bibinfo {volume} {68}},\ \bibinfo
  {pages} {085109} (\bibinfo {year} {2003})}\BibitemShut {NoStop}%
\end{thebibliography}%

\end{document}